\documentclass[prd,aps,preprint,nofootinbib]{revtex4}

\usepackage{graphicx}

\newcommand{\bear}{\begin{array}}  \newcommand{\eear}{\end{array}}
\newcommand{\bea}{\begin{eqnarray}}  \newcommand{\eea}{\end{eqnarray}}
\newcommand{\beq}{\begin{equation}}  \newcommand{\eeq}{\end{equation}}
\newcommand{\bef}{\begin{figure}}  \newcommand{\eef}{\end{figure}}
\newcommand{\bec}{\begin{center}}  \newcommand{\eec}{\end{center}}
  
\newcommand{\la}{\left\langle} \newcommand{\ra}{\right\rangle}

\def\lrf#1#2{ \left(\frac{#1}{#2}\right)}
\def\lrfp#1#2#3{ \left(\frac{#1}{#2}\right)^{#3}}

\def\lrf#1#2{ \left(\frac{#1}{#2}\right)}
\def\lrfp#1#2#3{ \left(\frac{#1}{#2}\right)^{#3}}

\preprint{IPMU 08-0032}

\begin{document}

\title{Isocurvature fluctuations in Affleck-Dine mechanism
 and constraints on inflation models}

\author{Shinta Kasuya $^a$, Masahiro Kawasaki $^{b,c}$, and Fuminobu
Takahashi $^c$}

\affiliation{
$^a$ Department of Information Science, 
     Kanagawa University, Kanagawa 259-1293, Japan\\
$^b$ Institute for Cosmic Ray Research,
     University of Tokyo, Chiba 277-8582, Japan\\
$^c$ Institute for the Physics and Mathematics of the Universe, 
     University of Tokyo, Chiba 277-8582, Japan}

\date{May, 2008}

\begin{abstract}
We reconsider the Affleck-Dine mechanism for baryogenesis and show
that the baryonic isocurvature fluctuations are generated in many
inflation models in supergravity. The inflationary scale and the
reheating temperature must satisfy certain constraints to avoid too
large baryonic isocurvature fluctuations.
\end{abstract}


\maketitle

\setcounter{footnote}{1}
\renewcommand{\thefootnote}{\fnsymbol{footnote}}

\section{Introduction}
\label{sec:1}

The origin of the baryon asymmetry in our universe remains a big
mystery in modern cosmology. The Affleck-Dine (AD) mechanism provides
one of the promising baryogenesis scenarios~\cite{AD}.  It utilizes a
flat direction of the supersymmetric standard model, which possesses a
non-zero baryon (or lepton) number.  A flat direction responsible for
the AD mechanism is referred to as the AD field.  The AD field is
assumed to develop a large expectation value during inflation, and it
starts to oscillate after inflation when the cosmic expansion rate
becomes comparable to its mass. The baryon number is effectively
created at the onset of the oscillations. Finally, the AD field decays
into the ordinary quarks, leaving the universe with a right amount of
the baryon asymmetry.

The scalar potential along the flat direction is crucial for the AD
mechanism sketched above to work.  The flat direction has a vanishing
potential at the level of the renormalizable operators in the limit of
supersymmetry (SUSY). In other words, the flat directions can be
lifted by the non-renormalizable operators and the SUSY breaking
effects.  During inflation, SUSY is largely broken by the inflaton
potential~\cite{DRT}. In particular, the radial component of the AD
field generically acquires a mass of the order of the Hubble
parameter, referred to as the Hubble-induced mass, due to supergravity
effects in the $F$-term inflation models.  The sign of the
Hubble-induced mass is assumed to be negative for the AD field to
develop a large field value during inflation.

The flat directions can be lifted also by non-renormalizable operators
in a superpotential. In fact, the non-renormalizable operator not only
lifts the potential at large scales, but also provides a baryon-number
violation needed to generate the baryon asymmetry.  Our concern here
is the strength of the baryon-number violation during and after
inflation.  It is often claimed that, during inflation, there appears
a baryon-number violating $A$-term with a coefficient comparable to the
Hubble parameter.  Such a large $A$-term is referred to as the
Hubble-induced $A$-term. If there were indeed the Hubble-induced
$A$-term during inflation, the phase component of the AD field would
acquire a mass of the order of the Hubble parameter, and hence would
quickly settle down in one of the minima given by the Hubble-induced
$A$-term. Thus, as long as the Hubble-induced $A$-term is not
suppressed, the phase component of the AD field does not have any
sizable fluctuations beyond the horizon scale during inflation.

In this paper we reconsider the AD mechanism and show that the
Hubble-induced $A$-term is suppressed in the most $F$-term inflation
models.  As a result, the phase direction of the AD field is rather
flat and therefore quantum fluctuations along that direction develop
during inflation. Those fluctuations turn into the baryonic
isocurvature fluctuations, which are now tightly constrained by the
observations of the cosmic microwave background (CMB).  Assuming that
the AD mechanism is responsible for generating the baryon asymmetry of
the universe, we will show that both the inflation scale and the
reheating temperature must satisfy a certain constraint.

Lastly, let us comment on the differences of the present paper from
the works in the past. The baryonic isocurvature perturbations in the
AD mechanism were discussed\footnote{
It was first pointed out in Ref.~\cite{Yokoyama:1993gb} that their amplitude could
be observably large without discussion of Hubble-induced mass and $A$ terms. }
for instance in Refs.~\cite{EM,KT}, both
of which focused on the $D$-term inflation
model~\cite{Halyo:1996pp}. This is partly because the absence of the
Hubble-induced $A$-term as well as the Hubble-induced mass term is
obvious in the $D$-term inflation.  It was also pointed out in
Ref.~\cite{EM} that the Hubble-induced $A$-term can be suppressed in a
certain class of the $F$-term inflation models. However, they did not
examine realistic inflation models such as the hybrid inflation
model~\cite{Copeland:1994vg}, and how generic the Hubble-induced $A$-term 
is suppressed was not clear. Indeed, it was often claimed that the presence of the
Hubble-induced $A$-term was unavoidable consequence of the $F$-term
inflation in supergravity. The purpose of this paper is to examine
representative inflation models and explicitly show that the
Hubble-induced $A$-term is suppressed, in addition 
to investigate its condition in general.  Furthermore, by using the
recent observational constraints on the isocurvature fluctuations, we
put tight bounds on the inflationary scale and the reheating
temperature for the first time.

\section{Affleck-Dine mechanism}
\label{sec:2}
Let us here briefly review the AD mechanism. In particular we explain
how the resultant baryon asymmetry is related to the baryon-number
violating operators, which will be important later on to estimate the
baryonic isocurvature fluctuations.

Flat directions are parameterized by composite gauge-invariant
monomial operators such as $udd$ or $LH_u$, and the dynamics of a flat
direction can be expressed in terms of a complex scalar field $\phi$.
The flat directions of the minimal supersymmetric standard model
are classified in Ref.~\cite{Gherghetta}. We assume that $\phi$
has a nonzero baryon number in the following.

First let us consider the scalar potential of $\phi$ in a flat space
time.  We assume that the AD field $\phi$ has a non-renormalizable
operator in the superpotential:
\beq
\label{nr}
W(\phi) \;=\; \frac{\lambda}{n} \phi^n,
\eeq
where $\lambda$ is a numerical coefficient, and $n$ is an integer, $n
= 4,5,6,7$ and $9$, which depends on flat directions.  We set
$\lambda$ to be real without loss of generality.  We adopt the Planck
unit, $M_P = 1$ ($M_P = 2.4 \times 10^{18}$\,GeV) here and in what
follows unless it is written explicitly. The above operator (\ref{nr})
lifts the flat direction at large scales.  In addition to the
non-renormalizable operator, the AD field has a soft SUSY breaking
mass, $m_\phi$. Thus the relevant potential for the AD field is
expressed as~\footnote{
In the gauge-mediated SUSY breaking models~\cite{GMSB}, the potential
coming from the SUSY breaking effects takes a different form and there
might be another contribution to the $A$-term. However, it does not
change the following arguments qualitatively.
}
\beq
V_0(\phi) \;=\; m_\phi^2 |\phi|^2 +\lambda^2 |\phi|^{2n-2} + V_A(\phi)
\eeq
with 
\beq
V_A(\phi) \;=\; a \lambda\, m_{3/2} \phi^n + {\rm h.c.}
\label{aterm}
\eeq
Here we add the $A$-term, $V_A$, where $a$ is a numerical
coefficient of order unity, and $m_{3/2}$ denotes the gravitino mass.
The presence of $V_A$ is generic, and one of the contributions comes
from the cross-term between (\ref{nr}) and the constant term in the
superpotential $W_0 \simeq m_{3/2}$.  The $A$-term explicitly violates
the baryon symmetry, which will play an important role in the AD
mechanism as described below.

 Let us make a comment that the $A$-term is proportional to the $R$
symmetry breaking, since one can assign the $R$ charge $2$ on the
operator (\ref{nr}), i.e., $R[\phi]=2/n$. Note that the $R$ symmetry
is necessarily violated by the constant term in the superpotential to
make the cosmological constant (almost) vanish.  Therefore the
$A$-term $V_A$ naturally arises in a flat spacetime, picking up the
$R$ symmetry breaking, $W_0$~\footnote{
This is actually the anomaly-mediated SUSY breaking
effect~\cite{AMSB}, since the operator (\ref{nr}) explicitly breaks
the scale invariance.
}. As we will discuss in the next section, however, it depends on the
inflation models whether there are larger $R$ breakings during and
after inflation.

When the inflaton dominates the energy density of the universe, SUSY
is largely broken by the potential energy of the inflaton, and the
potential of $\phi$ receives corrections.  We focus on the $F$-term
inflation models.  The scalar potential in supergravity is given by
\beq
V\;=\;e^{K}  \left(D_iW g^{i j^*} (D_j W)^* - 3|W|^2\right),
\label{supergravity-sp}
\eeq
where we adopt the usual convention that a subscript $i$ denotes a
derivative with respect to a scalar field $\phi_i$.  The sum over the
indices $i, j, \dots$ is understood unless otherwise stated.  $K$ is
the K\"ahler potential, $W$ is the superpotential, and $g^{i j^*}$ is
the inverse of the K\"ahler metric $g_{i j^*}$.  During inflation, the
inflaton potential is related to the Hubble expansion rate:
\beq
V(I)\;\simeq\; 3 H^2.
\eeq
If there is a quartic coupling between the AD field and the inflaton
in the K\"ahler potential, $K = c\, |\phi|^2|I|^2$ with $c > 1$, the
AD field has a negative Hubble-induced mass term.  The scalar
potential of $\phi$ can be written as
\beq
V(\phi) \;\simeq\; - c_H H^2 |\phi|^2 + V_0(\phi),
\label{vphi}
\eeq
where $c_H \equiv 3(c-1)$ is a numerical coefficient of order unity.
For simplicity we set $c_H = 1$ in the following. The expression
(\ref{vphi}) is actually valid as long as the inflaton dominates the
energy density of the universe, and we can use (\ref{vphi}) after
inflation until the reheating is completed.

There might be a further correction called the Hubble-induced
$A$-term, given by
\beq
V_{AH}(\phi) \;=\; b \,\lambda H \phi^n + {\rm h.c.},
\label{hia}
\eeq
which is similar to the $A$-term, $V_A$ (see (\ref{aterm})).  Here $b$
is a numerical coefficient, and its size is important for the baryonic
isocurvature fluctuations.  In this section we drop the Hubble-induced
$A$-term, and the condition for its appearance will be discussed
in detail in the next section.

Let us now describe the dynamics of the AD field. To this end we
decompose $\phi$ into the radial and phase components:
\beq
\phi \;=\;\frac{ \varphi}{\sqrt{2}} e^{i \theta}.
\eeq
When $H \gg m_\phi$,  the potential minimum of $\varphi$ is located at
\beq
\varphi_{\rm min} \; \simeq \alpha_n \lrfp{H}{\lambda}{\frac{1}{n-2}}
\label{eq:phir}
\eeq
with $\alpha_n \equiv (2^{n-2}/(n-1))^{1/2(n-2)}$. During inflation,
the $\varphi$ stays at the minimum, $\varphi = \varphi_{\rm min}$.
After inflation, the Hubble parameter decreases with time, and so does
the minimum.  The radial component $\varphi$ continues to track
$\varphi_{\rm min}$ until $H \sim m_\phi$, since its effective mass is
comparable to the velocity of the instantaneous minimum, i.e.,
$m_\phi^{\rm(eff)} = H \sim \dot{\varphi}_{\rm min}/\varphi_{\rm
min}$.  On the other hand, the phase component $\theta$ has a rather
flat potential coming only from $V_A$, and it is written as
\beq
V_A  \;=\; \frac{|a|}{\sqrt{n-1}} \,m_{3/2} H \varphi^2
\cos\left(n \theta + {\rm arg}[a]\right),\
\label{eq:aterm}
\eeq
where we have used $\varphi = \varphi_{\rm min}$.  The mass of
$\theta$ is of $O(\sqrt{m_{3/2} H})$, which is much smaller than $H$
before $\phi$ starts to oscillate~\footnote{ Precisely speaking, it is
$ \varphi \theta$ that has mass dimension one, rather than
$\theta$. Note also that the mass of $\varphi$, $m_\phi$, is
generically larger than or comparable to $m_{3/2}$. }.  Thus $\theta$
is in general deviated from the minima of $V_A$.

When the Hubble parameter becomes comparable to $m_\phi$, the AD field
starts to oscillate about the origin. The baryon asymmetry is
effectively generated at that moment, and the field is kicked in
the phase direction by the baryon-number-violating potential,
$V_{A}$. The baryon number density $n_B$ is defined by
\beq
n_B \;\equiv\; i \left(\phi \dot{\phi}^* - \phi^* \phi \right),
\eeq
where we assume that $\phi$ has an unit baryon number. The evolution of the baryon
number density is given by
\beq
\dot{n}_B + 3 H n_B = -i \phi \frac{\partial V_A}{\partial \phi} + {\rm h.c.},
\eeq
where the dot denotes a derivative with respect to the time.  The
baryon number density at the onset of the oscillations is estimated as
\beq
n_B \;\sim\; \frac{n}{m_\phi} 
|a| \lambda m_{3/2} \varphi_{osc}^n \sin\left(n \theta + {\rm arg}[a] \right),
\eeq
where $\varphi_{osc}$ represents $\varphi$ evaluated at $H = m_\phi$ by using (\ref{eq:phir}).
The baryon-to-entropy ratio is then given by
\beq
\frac{n_B}{s} \;\sim\; \frac{m_{3/2} }{m_\phi^2} \varphi_{osc}^2 \,T_{RH} 
\sin\left(n \theta + {\rm arg}[a] \right),
\label{nbs}
\eeq
where $T_{RH}$ is the reheating temperature, and we assume that
the reheating is not completed when the AD field starts the
oscillations, since otherwise too many gravitinos would be produced by
thermal scatterings.

From the arguments above, it is clear that the resultant baryon
asymmetry is directly related to the baryon-number violating operator,
$V_A$.  Our estimate on the baryon asymmetry actually remains
unchanged even if there is an unsuppressed Hubble-induced
$A$-term. However, whether the isocurvature fluctuations in the baryon
asymmetry is generated crucially depends on the presence of the
Hubble-induced $A$-term.  If it is unsuppressed, $\theta$ does not
have any sizable fluctuations beyond the Hubble horizon scale during
inflation. On the other hand, if the Hubble-induced $A$-term is
suppressed, $\theta$ acquires quantum fluctuations of $O(H)$ during
inflation, which will turn into the baryonic isocurvature
fluctuations.  Therefore, the strength of the baryon-number violation
is a crucial issue, and we will discuss possible origins of the
Hubble-induced $A$-term in the next section.

\section{Hubble-induced $A$-terms}
\label{sec:3}

In this section we discuss the conditions for the Hubble-induced
$A$-terms to arise and will see that, in general, they are suppressed
in the $F$-term inflation models.  The Hubble-induced $A$-term is
given by (\ref{hia}).  The size of the numerical coefficient $b$ is
important to estimate the baryonic isocurvature fluctuations. Assuming
that the radial component $\varphi$ tracks the instantaneous minimum
(\ref{eq:phir}), one can express $V_{AH}$ in terms of $\varphi$ and
$\theta$:
\beq
V_{AH}  \;=\; \frac{|b|}{\sqrt{n-1}} \, H^2 \varphi^2
\cos\left(n \theta + {\rm arg}[b]\right).
\eeq
Therefore, if $b$ is of $O(1)$, the phase component $\theta$ has a mass
comparable to the Hubble parameter. If this is the case, $\theta$
cannot have sizable fluctuations beyond the Hubble horizon scale, and
the baryonic isocurvature fluctuations are absent.

What kind of conditions are necessary for the Hubble-induced $A$-term
to arise? The $R$ symmetry must be largely broken during inflation in
order to have sizable Huble-induced $A$-terms, since they violate the
$R$ symmetry by $\Delta R = 2$, as we will see. If there were not for
any $R$-breaking terms other than the constant term $W_0 \sim m_{3/2}$
in the superpotential, the Hubble-induced $A$-term could not arise; we
would only have the ordinary $A$-term (\ref{eq:aterm}).

If the supergravity effects are negligible, the superpotential during
inflation can be approximated by
\beq
W \;\simeq\; v^2 I,
\label{eq:w}
\eeq
where $I$ is the inflaton, and the Hubble parameter during inflation
is $H \simeq |F_I|/\sqrt{3} \sim v^2$, where $F_I \simeq - (W_I)^*=-v^2$ is the
$F$-term of the inflaton.  In the inflation models with multiple
scalar fields, the superfield $I$ in (\ref{eq:w}) may not be the
inflaton which slow-rolls generating the adiabatic density
perturbations. For definiteness we call such a field $I$ that has a
large $F$-term during inflation as the inflaton.  Most inflation
models such as new, hybrid, and their variants fall in this category.
We can then naturally assign the $R$-charge of the inflaton as, 
$R[I] = +2$.  In order to
have the Hubble-induced $A$-term, the following operators should be
unsuppressed:~\cite{DRT,Aterm}~\footnote{
Another operator, $ {\cal O}_1' = \int d^4\theta\, I^\dagger \phi \phi
+ {\rm h.c.}$, can also generate the Hubble-induced $A$-term. Such
operator is only allowed for $H_u H_d$ and $L H_u$
directions. However, $H_u H_d$ direction cannot create any baryon
asymmetry of the universe, and the inflaton $I$ must violate the
lepton number for $L H_u$ direction.
}
\bea
\label{op1}
{\cal O}_1 &=& \int d^4\theta\, \left(I \cdot |\phi|^2 + {\rm h.c.}\right),\\
\label{op2}
{\cal O}_2 &=& \int d^2\theta\, I \cdot \frac{\lambda}{n} \phi^n + {\rm h.c.},
\eea
where both ${\cal O}_1$ and ${\cal O}_2$ violate the $R$ symmetry by
$\Delta R = 2$ \footnote{
One can obtain the operator ${\cal O}_2$ from ${\cal O}_1$,
by rescaling $\phi$ as $(1+I)\phi \rightarrow \phi$ in the presence of
the non-renormalizable operator (\ref{nr}).}.
The operator ${\cal O}_1$ induces a kinetic mixing, $g_{I \bar{\phi}} = - g^{\phi {\bar I}} = \phi$,
and  the Hubble-induced $A$-term arises
from  $W_\phi g^{\phi \bar{I}} (W_I)^*$
(see Eq.~(\ref{supergravity-sp})). In the meantime,  one can see that
${\cal O}_2$ generates the Hubble-induced $A$-term by noting $|F_I| \sim H$.

If the operators ${\cal O}_1$ and ${\cal O}_2$ are not suppressed, the
inflaton $I$ is effectively singlet; there is {\it a priori} no reason to
expect that the inflaton does not have any other operators violating the
$R$-symmetry.  This can cause two cosmological problems. One is that,
since $I$ can be treated as a singlet, it is hard to have a flat
potential for successful inflation without severe fine-tunings,
because all terms like $I, I^2, I^3, I^4, \ldots$, which otherwise
could be forbidden by symmetry, will appear in the inflaton potential.
Thus, a severe fine-tuning will be necessary in general.  The other is
that the inflaton generically couples to the SUSY breaking sector, so
decays into the gravitinos with an intolerably large
rate~\cite{Kawasaki:2006gs,Asaka:2006bv,Endo:2006tf,Endo:2006qk,Endo:2007ih}. It
is nothing but a cosmological disaster unless the gravitino mass is
extremely heavy or light.  Therefore, in order to evade these
problems, it is usually necessary to impose that the $R$-symmetry is a
(relatively) good symmetry during inflation, which leads to suppression of
both ${\cal O}_1$ and ${\cal O}_2$.

So far we have neglected the supergravity effects. In particular, the
field value of the inflaton $I$ is assumed to be smaller than the
Planck scale: $|I| \ll 1$.  This is the case for most inflation models
except for the so-called large scale inflation models. When the
supergravity effects are important, the Hubble-induced $A$-term is
generically generated if one of the following conditions is met
\bea
\label{condition1}
|I| &\gtrsim& 1,\\
\label{condition2}
|W(I)| &\sim& H.
\eea
If we substitute (\ref{condition1}) into (\ref{eq:w}), we obtain
 (\ref{condition2}). Of course, this naive argument may not be
 applicable when the full supergravity effects are taken into account:
 in principle, there can be cancellation between several contributions
 to $W$.  If $|I| \gtrsim 1$, one can obtain the operator ${\cal O}_1$
 with a coefficient $\la I \ra \gtrsim 1$, from the quartic coupling
 $K = c\, |\phi|^2|I|^2$ needed for the negative Hubble-induced mass
 term, after expanding the inflaton about its field value.  On the
 other hand, if $|W(I)| \sim H$, it gives rise to the Hubble-induced
 $A$-term as in the ordinary $A$-term, since $|W(I)|$ measures the
 breaking of the $R$-symmetry.

Our concern here is whether the conditions (\ref{condition1}) and
(\ref{condition2}) are satisfied in the realistic large-scale
inflation models such as the chaotic inflation model.  Although it is
hard to construct a successful chaotic inflation model in
supergravity, one was proposed in Ref.~\cite{KYY}. Actually, however,
neither condition is met in this model.  The superpotential is $W=m X
Y$, where $m \simeq 2\times 10^{13}{\rm\,GeV}$ is the inflaton mass.
$Y$ is charged under a shift symmetry and has a large expectation
value during inflation, ${\rm Im}[Y] \gtrsim1$.  On the other hand, it
is $X$ that has a large $F$-term, so we must regard $I=X$ in
Eqs.(\ref{condition1}) and (\ref{condition2}). During inflation, the
$X$ stays at the origin, $X=0$ \footnote{
In this respect, this chaotic inflation model can be classified into
the category considered below Eq.(\ref{eq:w}). In fact, $R[X]=+2$ and
$v^2 = m\langle I \rangle$.
}. Therefore neither (\ref{condition1}) nor (\ref{condition2}) is
satisfied.  Let us comment on other large-scale inflation models.  The
no-scale type chaotic inflation was constructed in
Ref.~\cite{MSYY}. Besides the problem as to the moduli stabilization
and too large gravitino mass in this model, the no-scale nature leads
to the vanishing Hubble-induced mass and $A$-terms. On the other hand, a
natural inflation model was constructed in Ref.{\cite{Kallosh}} with
the use of a shift symmetry.  In this model, Hubble-induced $A$-terms
can be obtained, but the gravitino mass is comparable to the Hubble
parameter during inflation. Therefore the Hubble-induced $A$-term is
more or less comparable to the ordinary $A$-term.  We do not consider
such cases, and focus on those inflation models that have much larger Hubble parameter 
 than the gravitino mass.

For completeness we consider whether the Hubble-induced $A$-term appears after
inflation, because its appearance after inflation will diminish
the amplitude of the fluctuations in the $\theta$-direction produced
during inflation.  The energy density of the universe after inflation
is dominated by the inflaton oscillating about its potential minimum
until it decays into radiation.  Expanding the inflaton about its
potential minimum, one can express the K\"ahler potential and the
superpotential respectively as
\bea
\label{kahler-after-inf}
K&=& |I|^2+ \cdots = I_{\rm min}^* {\hat I} + I_{\rm min} {\hat I}^{\dag} 
+ |{\hat I}|^2 + \cdots,\\
W &=& \frac{1}{2} M (I - I_{\rm min})^2 + \cdots = \frac{1}{2} M {\hat I}^2 + \cdots,
\eea
where $I_{\rm min}$ denotes the minimum of the potential, and ${\hat
I} \equiv I - I_{\rm min}$ is the excitation of the inflaton about
$I_{\rm min}$.  The Hubble parameter is related to the amplitude of
the inflaton as
\beq
H^2 \;\sim\; |F_I|^2 \sim M^2 \la {\hat I}^2\ra,
\eeq
where $F_I = -M {\hat I}^*$ denotes the $F$-term of the inflaton.
Since the inflaton oscillates around its potential minimum, the
averaged values of $F_I$ vanishes. Therefore the Hubble-induced
$A$-term is suppressed after inflation~\cite{Kamada:2008sv}.

We therefore conclude that it is hard to generate the Hubble-induced
$A$-terms in most, if not all, of the supergravity $F$-term inflation
without having severe fine-tunings or cosmological difficulties. As we
will see in the next section, the absence of the unsuppressed
Hubble-induced $A$-term leads to the baryonic isocurvature
fluctuations.

\section{Baryonic isocurvature fluctuations}
\label{sec:4}
If there is no sizable Hubble-induced $A$-term, the phase of the AD
field is effectively massless during inflation. Then it acquires
quantum fluctuations,
\beq
\delta \theta = \frac{H_{\rm inf}}{2 \pi \varphi_{\rm inf}},
\label{deltheta}
\eeq
where $\varphi_{\rm inf}$ denotes the field value of the $\varphi$
during inflation, obtained by substituting $H = H_{\rm inf}$ into
(\ref{eq:phir}).  We neglect the tilt of the fluctuations and assume
that the radial component does not have fluctuations beyond the
horizon scale.  Let us define the baryonic isocurvature fluctuation
$S_{b\gamma}$ as
\beq
S_{b\gamma} \;\equiv\; \frac{\delta \rho_B}{\rho_B} - 
\frac{3}{4}\frac{\delta \rho_\gamma}{\rho_\gamma} =
\delta\log \lrf{n_B}{s},
\label{Sbgamma}
\eeq
where $\rho_B$ and $\rho_\gamma$ denote the energy densities of the
baryons and photons, respectively, and we use $\rho_B \propto
n_B$ and $\rho_\gamma^{3/4} \propto s$ in the second equality.  Inserting
Eq.~(\ref{nbs}) into Eq.~(\ref{Sbgamma}), we obtain
\beq
S_{b\gamma}\;=\; n \cot\left(n \theta_{\rm inf} + {\rm arg}[a] \right) \delta \theta.
\label{biso}
\eeq

The observations of CMB have shown that the density perturbations are
predominantly adiabatic, and therefore the isocurvature perturbations
are tightly
constrained~\cite{Bean:2006qz,Trotta:2006ww,Keskitalo:2006qv,Sekiguchi}.
The latest WMAP 5yr data puts an upper bound at $95\%$
C.L.~\cite{Komatsu:2008hk},
\beq
\left|\frac{\Omega_b}{\Omega_c} S_{b \gamma}\right| \;\lesssim\; 
\left(\frac{0.067}{1-0.067} \cdot 2.4 \times 10^{-9}\right)^\frac{1}{2} \simeq 1.3 \times 10^{-5},
\eeq
where the isocurvature fluctuation is taken to be uncorrelated with
the adiabatic one, and $\Omega_b$ and $\Omega_c$ denote the density
parameters of the baryon and the cold dark matter, respectively. With
$\Omega_b \simeq 0.046$ and $\Omega_c \simeq
0.23$~\cite{Komatsu:2008hk}, we arrive at
\beq
|S_{b\gamma}| \;\lesssim\;6.6 \times 10^{-5}.
\label{wmap}
\eeq

Using (\ref{nbs}), (\ref{deltheta}), (\ref{biso}) and (\ref{wmap}), we
can rewrite the constraint as
\beq
T_{RH} \;\lesssim\; 1.7\times 10^{-7} \,\frac{1}{n^2}
 \frac{M_P^2}{m_{3/2}} \left(\frac{m_{\phi}}{H_{inf}}\right)^{\frac{2n-6}{n-2}}  \lrf{n_B}{s} \, \Theta,
\label{const}
\eeq
with
\beq
\Theta \equiv \frac{\sin\left(n \theta_{\rm inf} + {\rm arg}[a]\right)}{\cos^2\left(n \theta_{\rm inf} + {\rm arg}[a]\right)},
\eeq
where we eliminate $\varphi_{\rm inf}$ by using $\varphi_{\rm inf}
\simeq (H_{\rm inf}/m_\phi)^{1/(n-2)}\, \varphi_{\rm osc}$.  This is
the main result of this paper.

\begin{figure}[t]
\includegraphics[width=16cm]{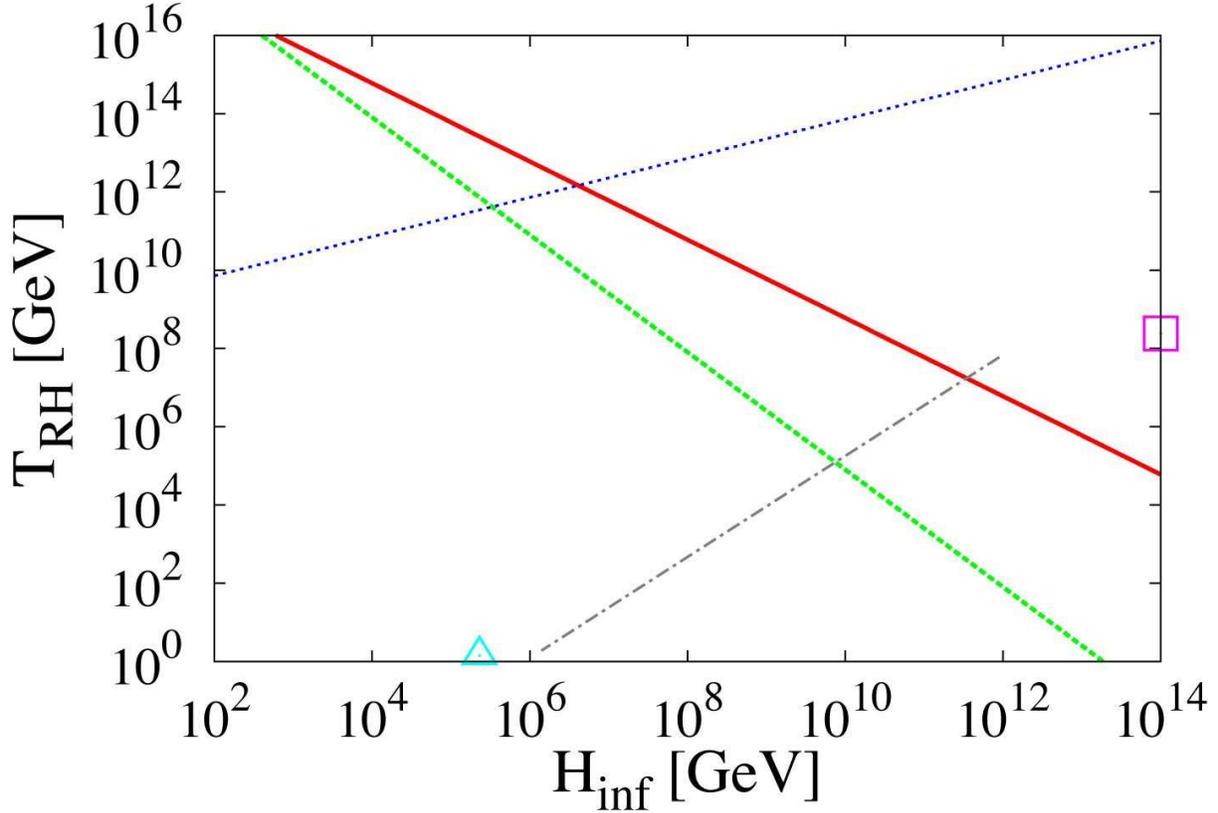}
\caption{Upper bounds on inflation models from AD isocurvature
fluctuations for $n=4$ (red, solid lines) and 6 (green, dashed
lines). We set $\Theta=1$ and $m_{3/2} = 1$ TeV.  
Also shown are the bounds of the reheating
temperature. Blue (dotted) line is the model-independent global upper
limit, while lower limits are shown for the chaotic, new, and hybrid inflations 
by pink square, light blue triangle, and gray dot-dashed line,
respectively.}
\label{fig1}
\end{figure}

The constraint (\ref{const}) reads
\begin{equation}
\label{n4}
  T_{RH} \;\lesssim\; 6 \times 10^{6} \,\textrm{GeV} \,
\lrfp{m_{3/2}}{1{\rm TeV}}{-1} \lrf{m_\phi}{1{\rm TeV}} \lrfp{H_{\rm inf}}{10^{12}{\rm GeV}}{-1} \Theta,
\end{equation}
for $n=4$, and
\begin{equation}
\label{n6}
  T_{RH} \;\lesssim\; 80 \,\textrm{GeV} \,
\lrfp{m_{3/2}}{1{\rm TeV}}{-1} \lrfp{m_\phi}{1{\rm TeV}}{\frac{3}{2}}
 \lrfp{H_{\rm inf}}{10^{12}{\rm GeV}}{-\frac{3}{2}} \Theta,
\end{equation}
for $n=6$, where we use $n_B/s = 8.8\times 10^{-11}$.  We show
the constraints (\ref{n4}) and (\ref{n6}) as red (solid) and greed
(dashed) lines,respectively, in Fig.~\ref{fig1}.  One can see that the
upper bound on the reheating temperature becomes severer for larger
$H_{\rm inf}$ and $n$.

A comment on Q-ball formation follows. It is known that
non-topological solitons called Q balls may be produced associated
with AD mechanism~\cite{Coleman,Qball}. 
As long as those Q balls decay
eventually into the ordinary quarks, the estimates of baryon number
and isocurvature fluctuations remain intact.
 This is the case of
the gravity- or anomaly-mediation. In the gauge-mediation, the
amplitude of $A$-terms and the dynamics of the field is more
model-dependent.  Also the Q balls can be stable, and only a small
fraction of the baryon number may be released from
them~\cite{Qevap}. Then the constraint on the reheating temperature
will be relaxed by a factor $\Delta^{-(n-2)/(2n-6)}$, where $\Delta$
denotes the fraction of the baryon number emitted from the Q ball.
Note however that the lightest supersymmetric particle produced
by the decay of the Q balls, or the  Q balls themselves if stable, 
may contribute to the dark matter abundance.
In this case, the constraint (\ref{const}) may become severer,
since the constraint on the dark matter isocurvature perturbations
are severer than that on the baryonic one.

\section{Constraints on inflation models}
\label{sec:5}
There is an absolute upper bound on the reheating temperature obtained
by $\rho_{rad}(T_{RH}) \le 3 H_{\rm inf}^2 M_P^2$,
\begin{eqnarray}
T_{RH} & \le & \left(\frac{90}{\pi^2 g_*}\right)^{\frac{1}{4}} \sqrt{H_{inf} M_P} \nonumber \\
& \sim & 7.2 \times 10^{14} \,{\rm GeV}\left(\frac{g_*}{200}\right)^{-\frac{1}{4}}  
\left(\frac{H_{inf}}{10^{12} \textrm{ GeV}}\right)^{\frac{1}{2}},
\label{upper}
\end{eqnarray}
On the other hand, the inflaton inevitably decays into the
standard-model sector through the top Yukawa coupling, if it has
non-vanishing expectation value at the potential minimum, i.e., if
$I_{\rm min} \ne 0$~\cite{Endo:2006qk}. The lower bound reads
\begin{eqnarray}
T_{RH} & \gtrsim & 1.9\times 10^3 {\rm GeV}\, |Y_t|\, \left(\frac{g_*}{200}\right)^{-\frac{1}{4}}  
 \left(\frac{I_{\rm min}}{10^{15} \textrm{ GeV}}\right) 
\left(\frac{m_I}{10^{12} \textrm{ GeV}}\right)^{\frac{3}{2}},
\label{topY}
\end{eqnarray}
where $Y_t$ is the top Yukawa coupling.  The reason why such a decay
proceeds is as follows.  When $I_{\rm min} \ne 0$, the inflaton has a
linear term in the K\"ahler potential as in
Eq.~(\ref{kahler-after-inf}). Then the inflaton ${\hat I}$ couples to
any particles that appear in the superpotential. This is because it is
a combination given by $e^{K/2}W$ that is more important in
supergravity at tree level, instead of $K$ and $W$. In other words,
using the K\"ahler transformation, one can remove the linear term of
the inflaton from the K\"ahler potential, and obtain a new
interaction, $W = (I_{\rm min}^*) {\hat I}\cdot Y_t\, T Q H_u$, in the
superpotential.

The reheating temperature is not known, due to our ignorance on the
inflation and the reheating processes; it can {\it a priori}
take any values between (\ref{topY}) and (\ref{upper}). The upper
bound on $T_{RH}$, (\ref{const}), therefore excludes large parameter
space of the reheating temperature, which are otherwise allowed~\footnote{
There are other constraints on the reheating temperature, coming from,
e.g., the thermal production of the gravitinos. Our constraint
(\ref{const}) is independent of those constraints. In fact,
(\ref{const}) gives a very stringent upper bound as (\ref{n6}), which
cannot be obtained otherwise.
}.  Since the bounds (\ref{topY}) and (\ref{upper}) depend on the
inflaton parameters such as the inflation scale, the mass and the field value at the
potential minimum, let us consider the bounds for representative
inflation models such as chaotic \cite{KYY}, new \cite{new}, and
hybrid \cite{hybrid} inflation models, separately.  The details of
each model are given in Appendix.

For the chaotic inflation model without a $Z_2$ symmetry, we expect
$I_{\rm min} \sim 1$. The inflaton mass is determined to be $m_I
\simeq 2 \times 10^{13}$ GeV, leading to $T_{RH} \gtrsim 2.4\times
10^8$ GeV from (\ref{topY}).  As can be seen in Fig.~\ref{fig1}, the
chaotic inflation is inconsistent with the constraint from the
baryonic isocurvature fluctuations, unless we impose a $Z_2$ symmetry
on the inflaton to suppress its expectation value as $|I_{\rm min}|
\ll 1$~\cite{Kawasaki:2006gs}. Note that the constraint is actually severer
in the chaotic inflation, 
since the tensor modes will leave less room for the baryonic isocurvature
fluctuations.

For the new inflation model, the inflaton parameters are $m_I \simeq 4\times 10^9$ GeV 
and $I_{\rm min} \simeq 3\times 10^{15}$ GeV for $m_{3/2}= 1$ TeV.
The lower bound on the reheating temperature is $T_{RH} \gtrsim 1.4$ GeV, 
which can satisfy the bound (\ref{const}). The inflation scale is relatively low, given by 
$H_{inf} \simeq {\cal O}(10^5)$ GeV.

For the hybrid inflation model, for $\kappa \sim 10^{-1} - 10^{-5}$, we
have $m_I \sim 10^{15} - 10^{10}$ GeV and $I_{\rm min} \simeq 10^{15}$
GeV for $H_{inf} \sim 10^{11} - 10^6$ GeV.  The lower bounds on the
reheating temperature read $T_{RH} \gtrsim 6.0 \times 10^7 - 1.9$
GeV. Therefore, large values of $\kappa$ are excluded by the bounds
for isocurvature fluctuations.

\section{Conclusion}
We have reconsidered the Affleck-Dine mechanism and discussed the
condition for the Hubble-induced $A$-term to arise.  It has
turned out that the $R$-symmetry needs to be largely broken 
during inflation. If this is the case, however, theoretical and cosmological
difficulties arise; one needs fine-tunings to make the inflaton
potential flat enough, and moreover, the inflaton decay produces too
many gravitinos.  Therefore, 
$R$-symmetry, under which the inflaton is charged, should be
a good symmetry to some extent, which results in the suppression 
of the Hubble-induced $A$-term.  Then the phase direction 
of the AD field is effectively massless during inflation, and its fluctuations contribute
to sizable baryonic isocurvature perturbations, while the fluctuation
in the radial direction is absent due to the Hubble-induced mass term
in the $F$-term inflation models. Note that the Hubble-induced $A$-term is
generally suppressed after inflation.
Using the latest WMAP-5yr data, we have derived tight constraints on
the inflation models in terms of the Hubble parameter during inflation
and the reheating temperature, based on the assumption that the AD
mechanism is responsible for the baryon asymmetry of the universe.
The constraint on the reheating temperature is very tight
especially for the high-scale inflation models and higher values of $n$.
We conclude that large-scale inflation models with high reheating temperatures 
are not preferred in the context of the AD mechanism.

\section*{Acknowledgments}
The work is supported by Grant-in-Aid for Scientific Research from
the Ministry of Education, Science, Sports, and Culture (MEXT), Japan, 
No.~17740156 (S.K.) and No. 14102004 (M.K.). This work was supported 
by World Premier International Research Center InitiativeiWPI Initiative), MEXT, Japan. 
This work was also supported
in part by JSPS-AF Japan-Finland Bilateral Core Program (S.K. and  M.K.).


\appendix
\section{Inflation models}
In the appendix, we give brief explanations on the inflation models we have considered
in the text.

\subsection{Chaotic inflation}
The K\"ahler potential is expressed as \cite{KYY}
\begin{equation}
K(I,I^\dagger) = \frac{1}{2} (I+I^\dagger)^2,
\end{equation}
which is invariant under the shift, $I \rightarrow I + i A$, where $A$
is a dimensionless real parameter. The inflaton field is identified as
the imaginary part of $I$, i.e., $\sqrt{2} \textrm{ Im} I$.  The
superpotential is written as
\begin{equation}
W(I,X) = m X I,
\end{equation}
which breaks the shift symmetry a bit with a small mass scale $m \simeq 2\times 10^{13}$ GeV ($\ll 1$).
Here another chiral multiplet $X$ is introduced, whose $F$-term is responsible for the potential
energy during and after inflation: $V \simeq |F_X|^2$.

\subsection{New inflation}
The K\"ahler potential and superpotential of the inflaton sector are
written as \cite{new}
\begin{eqnarray}
& & K(I,I^\dagger) = |I|^2+\frac{k}{4}|I|^4, \\
& & W(I) = v^2 I -\frac{g}{s+1}I^{s+1},
\end{eqnarray}
where $k$ and $g$ are constants, $s$ is an integer, and $v$ is the inflation energy scale.
In order to explain the observed density fluctuations, we need $v = 4\times 10^{-7} (g/0.1)^{-1/2}$
and $k \lesssim 0.03$ for $s=4$. After inflation, the vev of the inflaton becomes 
$I_{\rm min} \simeq (v^2/g)^{1/s}$, and its mass is given by $m_I \simeq s v^2/I_{\rm min}$.
The gravitino mass in this model is related to $v$ as 
$m_{3/2} \simeq s v^2 I_{\rm min} / (s+1)$, because the inflaton induces a spotaneous 
R-symmetry breaking.

\subsection{Hybrid inflation}
The superpotential of the inflaton sector is written as \cite{hybrid}
\begin{equation}
W(I,\psi,\bar{\psi}) = I ( v^2 - \kappa \psi \bar{\psi} ),
\end{equation}
where the two superfields $\psi(+1)$ and $\bar{\psi}(-1)$ are the waterfall fields, which are charged
under a $U(1)$ gauge symmetry. $\kappa$ is a coupling constant
and $v$ is the inflation energy scale. To account for the WMAP normalization on the
density fluctuations, $v$ and $\kappa$ are 
related as $v \simeq 2\times 10^{-3} \kappa^{1/2}$ for $\kappa \gtrsim 10^{-3}$, while
$v \simeq 2\times 10^{-2} \kappa^{5/6}$ for $\kappa \lesssim 10^{-3}$.



\begin{thebibliography}{90}


\bibitem{AD}
I.~Affleck and M.~Dine,
Nucl.\ Phys.\ B {\bf 249}, 361 (1985).



\bibitem{DRT}
M.~Dine, L.~Randall and S.~Thomas,
Nucl.\ Phys.\ B {\bf 458}, 291 (1996).

\bibitem{Yokoyama:1993gb}
  J.~Yokoyama,
  Astropart.\ Phys.\  {\bf 2}, 291 (1994).
 

 
\bibitem{EM}
  K.~Enqvist and J.~McDonald,
  Phys.\ Rev.\ Lett.\  {\bf 83}, 2510 (1999);
  Phys.\ Rev.\  D {\bf 62}, 043502 (2000).
  
 \bibitem{KT}
  M.~Kawasaki and F.~Takahashi,
  Phys.\ Lett.\  B {\bf 516}, 388 (2001).
  
  
\bibitem{Halyo:1996pp}
  E.~Halyo,
  Phys.\ Lett.\  B {\bf 387}, 43 (1996);
  P.~Binetruy and G.~R.~Dvali,
  Phys.\ Lett.\  B {\bf 388}, 241 (1996).

\bibitem{Copeland:1994vg}
  E.~J.~Copeland, A.~R.~Liddle, D.~H.~Lyth, E.~D.~Stewart and D.~Wands,
  Phys.\ Rev.\ D {\bf 49}, 6410 (1994);
  G.~R.~Dvali, Q.~Shafi and R.~K.~Schaefer,
  Phys.\ Rev.\ Lett.\  {\bf 73}, 1886 (1994);
  A.~D.~Linde and A.~Riotto,
  Phys.\ Rev.\ D {\bf 56}, 1841 (1997).
 
  
  
\bibitem{Gherghetta}
  T.~Gherghetta, C.~F.~Kolda and S.~P.~Martin,
  Nucl.\ Phys.\  B {\bf 468}, 37 (1996).
  
 \bibitem{GMSB}
  M.~Dine, A.~E.~Nelson and Y.~Shirman,
  Phys.\ Rev.\ D {\bf 51}, 1362 (1995);
  M.~Dine, A.~E.~Nelson, Y.~Nir and Y.~Shirman,
  Phys.\ Rev.\ D {\bf 53}, 2658 (1996);
  For a review, see, for example, 
  G.~F.~Giudice and R.~Rattazzi,
  Phys.\ Rep.\  {\bf 322}, 419 (1999),
  and references therein.
  
 
   \bibitem{AMSB} 
  L.~Randall and R.~Sundrum,
  Nucl.\ Phys.\ B {\bf 557}, 79 (1999);
  G.~F.~Giudice, M.~A.~Luty, H.~Murayama and R.~Rattazzi,
  JHEP {\bf 9812}, 027 (1998);
  J.~A.~Bagger, T.~Moroi and E.~Poppitz,
  JHEP {\bf 0004}, 009 (2000).
  
  

  
 \bibitem{Aterm}
  S.~Kasuya and M.~Kawasaki,
  Phys.\ Rev.\  D {\bf 74}, 063507 (2006);
  S.~Kasuya,
  J.\ Phys.\ A  {\bf 40}, 6999 (2007)
  [arXiv:hep-ph/0610428].

\bibitem{Kawasaki:2006gs}
  M.~Kawasaki, F.~Takahashi and T.~T.~Yanagida,
  Phys.\ Lett.\  B {\bf 638}, 8 (2006);
  Phys.\ Rev.\  D {\bf 74}, 043519 (2006).
  
\bibitem{Asaka:2006bv}
  T.~Asaka, S.~Nakamura and M.~Yamaguchi,
  Phys.\ Rev.\  D {\bf 74}, 023520 (2006).
  
  
\bibitem{Endo:2006tf}
  M.~Endo, K.~Hamaguchi and F.~Takahashi,
  Phys.\ Rev.\  D {\bf 74}, 023531 (2006).
  
\bibitem{Endo:2006qk}
  M.~Endo, M.~Kawasaki, F.~Takahashi and T.~T.~Yanagida,
  Phys.\ Lett.\  B {\bf 642}, 518 (2006).
  
\bibitem{Endo:2007ih}
  M.~Endo, F.~Takahashi and T.~T.~Yanagida,
  Phys.\ Lett.\  B {\bf 658}, 236 (2008);
  Phys.\ Rev.\  D {\bf 76}, 083509 (2007).


 

\bibitem{KYY}
  M.~Kawasaki, M.~Yamaguchi and T.~Yanagida,
  Phys.\ Rev.\ Lett.\  {\bf 85}, 3572 (2000);
  Phys.\ Rev.\  D {\bf 63}, 103514 (2001).


\bibitem{MSYY}
  H.~Murayama, H.~Suzuki, T.~Yanagida and J.~Yokoyama,
  Phys.\ Rev.\  D {\bf 50}, 2356 (1994).


\bibitem{Kallosh}
  R.~Kallosh,
  Lect.\ Notes Phys.\  {\bf 738}, 119 (2008)
  [arXiv:hep-th/0702059].
  

\bibitem{Kamada:2008sv}
  K.~Kamada and J.~Yokoyama,
  arXiv:0803.3146 [hep-ph].
  
  
\bibitem{Bean:2006qz}
  R.~Bean, J.~Dunkley and E.~Pierpaoli,
  Phys.\ Rev.\  D {\bf 74}, 063503 (2006).

\bibitem{Trotta:2006ww}
  R.~Trotta,
  Mon.\ Not.\ Roy.\ Astron.\ Soc.\ Lett.\  {\bf 375}, L26 (2007).

\bibitem{Keskitalo:2006qv}
R.~Keskitalo, H.~Kurki-Suonio, V.~Muhonen and J.~Valiviita,
  JCAP {\bf 0709}, 008 (2007).

\bibitem{Sekiguchi}
  M.~Kawasaki and T.~Sekiguchi,
  arXiv:0705.2853 [astro-ph].
  
\bibitem{Komatsu:2008hk}
  E.~Komatsu {\it et al.}  [WMAP Collaboration],
  arXiv:0803.0547 [astro-ph].

\bibitem{Coleman}
S.~R.~Coleman,
Nucl.\ Phys.\ B {\bf 262}, 263 (1985)
[Erratum-ibid.\ B {\bf 269}, 744 (1986)].


\bibitem{Qball}
A.~Kusenko and M.~E.~Shaposhnikov,
Phys.\ Lett.\ B {\bf 418}, 46 (1998);
  K.~Enqvist and J.~McDonald,
  Phys.\ Lett.\  B {\bf 425}, 309 (1998);
  Nucl.\ Phys.\  B {\bf 538}, 321 (1999);
  S.~Kasuya and M.~Kawasaki,
 Phys.\ Rev.\ D {\bf 61}, 041301(R) (2000).

\bibitem{Qevap}
M.~Laine and M.~E.~Shaposhnikov,
  Nucl.\ Phys.\  B {\bf 532}, 376 (1998);
  R.~Banerjee and K.~Jedamzik,
  Phys.\ Lett.\  B {\bf 484}, 278 (2000);
  S.~Kasuya and M.~Kawasaki,
  Phys.\ Rev.\ D {\bf 64}, 123515 (2001).



\bibitem{new}
K.~I.~Izawa and T.~Yanagida,
  Phys.\ Lett.\  B {\bf 393}, 331 (1997).


\bibitem{hybrid}
E.~J.~Copeland, A.~R.~Liddle, D.~H.~Lyth, E.~D.~Stewart and D.~Wands,
  Phys.\ Rev.\  D {\bf 49}, 6410 (1994);
G.~R.~Dvali, Q.~Shafi and R.~K.~Schaefer,
  Phys.\ Rev.\ Lett.\  {\bf 73}, 1886 (1994);
A.~D.~Linde and A.~Riotto,
  Phys.\ Rev.\  D {\bf 56}, 1841 (1997).
  
\end{thebibliography}
\end{document}